# A Bayesian Method for Estimating Uncertainty in Excavated Material

Mehala Balamurali

Australian Centre for Field Robotics, The University of Sydney
mehala.balamurali@sydney.edu.au
http://orcid.org/0000-0003-0083-5772

**Abstract**

This paper proposes a method to probabilistically quantify the moments (mean and variance) of excavated material during excavation by aggregating the prior moments of the grade blocks around the given bucket dig location. By modelling the moments as random probability density functions (pdf) at sampled locations, a formulation of the sums of Gaussian based uncertainty estimation is presented that jointly estimates the location pdfs, as well as the prior values for uncertainty coming from ore body knowledge (obk) sub block models. The moments calculated at each random location is a single Gaussian and they are the components of Gaussian mixture distribution. The overall uncertainty of the excavated material at the given bucket location is represented by the Gaussian Mixture Model (GMM) and therefore moment matching method is proposed to estimate the moments of the reduced GMM. The method was tested in a region at a Pilbara iron ore deposit situated in the Brockman Iron Formation of the Hamersley Province, Western Australia, and suggests a framework to quantify the uncertainty in the excavated material that hasn't been studied anywhere in the literature yet.

## 1. Introduction

Significant effort has put into characterise the spatial uncertainty of grades through ore control block models. Although, these geological models allow for reasonable uncertainty information to be attached to specific volumes, there is little, and no attention given to how these estimated values are further blend during excavation process. Uncertainty information behaves in non-trivial ways when material is being mixed. Not only the grade blocks are getting blended during excavation, also the process increase the uncertainty of the measurement of exact dig locations (Fig1). As a first step in the material tracking pipeline, estimating the exact material in the bucket and its uncertainty, can be utilised throughout the material movement pipeline.

In this paper, we propose a method to probabilistically quantify the moments of excavated material by aggregating the prior distribution of iron weight percentage (Fe wt%) of the grade blocks around the given bucket dig location. The model captures the uncertainty introduced by the uncertain bucket dig location, by stochastically simulating multiple bucket locations across the bench and then estimates the pdf of excavated material at each location. The final GMM constructed through a linear combination of Gaussian densities at each simulated random bucket dig locations gives the material uncertainty associated with the given bucket. The moments of the excavated material is then estimated using the method of moment matching. A case study was conducted in a test region at Pilbara iron ore deposit situated in the Brockman Iron Formation of the Hamersley Province, Western Australia.

The remainder of the paper is structured as follows. The need of the proposed work is presented under background at Section2. The detail about the uncertainty attached to the excavated material problem is presented in Section 3 and describes the mathematical description of the solution and the application of GMM-moment matching to estimate the moments. The Data and the results are presented in Sections 4 and 5 and are followed by the conclusions.

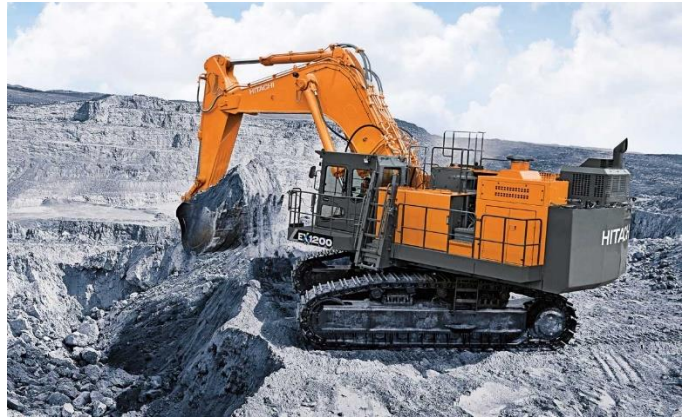

*Figure1: Hitachi, an ultra-large hydraulic excavator in Australia during trial excavation, July 2020, published in News and Events, Autonomous tech news*

## 2. Background

Accuracy of grade control determines the success of mine that directly influences the profitability of the overall mine operation. This begins with the quality of the mining program that controls how the ore is collected, stored and assayed geological data. This is then followed by the accurate prediction of grade variable of interest. These forecasted grades then determine the destination for waste and ore. Once the locations are determined, the material is moved using diggers and haulage trucks. However, tracking material only by its most likely type or mean grade results does not convey the confidence into the assumed values and their likelihood of deviating from the expectation. Once the statistical probability distribution of the variable is known or can be assumed, it is possible to derive confidence limits to describe the region within which the true value of the variable may be found.

The standard mine modelling practice often involves investing significant effort into the interpretation of the deposit and identification of 3D geological models to provide the block estimates. Irregular geometric shapes, grade blocks, are used to simplify the representative ore and waste regions. These grade blocks are further subdivided into smaller cuboid sizes in order to better model boundaries and the values are inferred in to the blocks using various geostatistical simulation methods (Rossi & Deutsch, 2014, Isaaks & Srivastava, 1989, Srivastava, 1987; Boucher and Dimitrakopoulos, 2012, Melkumyan and Ramos, 2009).

However, in practice it is difficult to precisely define the ore-waste boundaries. Thus this introduces dilution and ore loss (Dimitrakopoulos & Godoy, 2014; Verly, 2005). The choice of block size becomes inadequate to precisely and feasibly model the grade estimation and its uncertainty, because, the mining equipment cannot exactly stop digging at the boundaries and it is impossible to mine isolated ore or waste blocks. The infeasibility of freely mining block by block is the major impediment of the current grade control methodologies. Some studies proposed optimal dig-limits by taking into account the digability of excavating equipment. Digability measures the difficulty of mining an ore or waste region depends on the geometry of the region to be excavated (Isaaks et. al 2014, Norrena and Deutsch, 2001). In order to better create the selective mining unit Wilde and Deutsch (2007) replaced dig-limits with a truck-by-truck approach where the mine block is further discretised ( Vasylchuk and Deutsch, 2015). However, there are only very few studies available in this direction and the methods are only available in publication. Furthermore, the data used in these studies were synthesised on grid spaces. While the grade control model blocks are vertical, the excavator digs a face at the rill angle. This also highly depends on the operator's skill and operating conditions. Thus, diggers further blend the material when they reclaim the material from the ground and introduce high uncertainty when they load material to trucks. Hence the geological domain developed for grade estimation do not always meet the production target.

The key idea of this paper is to take the next step forward from geostatistical modelling of grades and block-by-block decision making using other information coming from the mine site. GPS positions of excavating equipment and excavators' bucket dig positional data (the position where the bucket engage with ground during excavating) are some of the available data sources at the time of digging. This paper presents a model approach to quantify the estimated mean grade for each bucket and the uncertainty of that estimation by utilising the excavator's bucket dig locations and parametric information coming from ore body knowledge.

A simple approach to estimate the values for a bucket, is to directly assign the pre-estimated moments of the 3D block to that bucket's content where the bucket falls or assign the sum of weighted values of the blocks that are intersecting with the corresponding bucket. The weights are the fraction of bucket volumes that intersects with the block. The uncertainty attached to the bucket content comes from the variance of spatially correlated blocks. However, bucket positions are not always sensed or measured on the stable-solid ground, because the process of excavation blends the material across the bench. Hence, the bucket dig locations shouldn't be constrained with respect to geo statistical grade blocks' boundaries. Instead, the material associated with the bucket dig location, can represent a mixture of material from the nearest sub-blocks.

In order to deal with the uncertain bucket dig locations, the proposed model uses the grade values taken from adjacent locations for the estimation of moments of excavated material. GMM and moment matching are the basis of the analytical technique proposed in this work.

### 3. Problem Statement and formulation

We are interested to model the probabilistic estimation p(Fs) of the Fe wt% of the material excavated by the single bucket based on blocks around the given bucket dig location x. We first defined a set of individual estimation, $Fs^j$ at random bucket dig location $x_j$, as independent events with respect to each other.

The system's uncertain variables are $\{v_{si}, F_{bi}\}$, where each $F_{bi}$ is an Fe wt% of block i. The prior distribution p($F_{bi}$) of the Fe wt% is known for block *i* from the ore body block model. The adjacent blocks' estimations are spatially correlated. The bucket content, for a sampled bucket dig location, is according to the volume intersection with relevant blocks as presented in Figure 2. Therefore, we model each $v_{si}$ as the intersecting volume of the bucket and the i[th] block, and assume this volume is perfectly known if the location of the bucket is known. That is, for a given bucket dig location $x_j$, the volumes $\{v_{s1}^j, \dots, v_{si}^j, \dots\}$ are known; where the bucket volume $v = \sum_i v_{si}^j$. Therefore, the excavated material at j[th] location is given by the model

$$F_s^j = \sum_i v_{si}^j F_{bi} \qquad \text{Eq1}$$

In Eq1, the conditional probability $p(F_s^j | x_j)$ is Gaussian, and so a joint probability p(Fs, x) can be obtained by generating samples $x_j \sim p(x)$, and computing the Gaussian $p(F_s^j | x_j)$ for each sample.

However, observations coming from geo spatial references should not be treated statistically independent. Those observations are highly correlated at the nearest locations. In the Eq1

$$F_s^j = v_{s1}^j F_{b1} + v_{s2}^j F_{b2} + \cdots + v_{sn}^j F_{bn}$$

If $(F_{b1}, F_{b2}, \ldots, F_{bn})$ are correlated then the joint distribution $p(F_{b1}, F_{b21}, \ldots, F_{bn})$ with moments,

$$\begin{bmatrix} \hat{F}_{b1} \\ \hat{F}_{b2} \\ \vdots \\ \hat{F}_{bn} \end{bmatrix}, \begin{bmatrix} \epsilon_{11} & \cdots & \epsilon_{1n} \\ \epsilon_{21} & \cdots & \epsilon_{2n} \\ \vdots & \ddots & \vdots \\ \epsilon_{n1} & \cdots & \epsilon_{nn} \end{bmatrix}$$

Where $\hat{F}_{b1}$ are the mean of block i and $\epsilon_{ij}$ are the correlation between the blocks i and j. If the bucket intersects with n blocks, then the weight matrix consists a covariance matrix of size $n \times n$. We define our model as

$$F_s = [v_{s1}, v_{s2}, \ldots, v_{sn}] \begin{bmatrix} F_{b1} \\ F_{b2} \\ \vdots \\ F_{bn} \end{bmatrix} \quad \text{Eq2}$$

Hence the mean at location $x_j$ is translated as

$$\hat{F}_s^j = [v_{s1}^j, v_{s2}^j, \ldots, v_{sn}^j] \begin{bmatrix} \hat{F}_{b1} \\ \hat{F}_{b2} \\ \vdots \\ \hat{F}_{bn} \end{bmatrix} \quad \text{Eq3}$$

and the covariance as

$$\epsilon_s^j = [v_{s1}^j, v_{s2}^j, \ldots, v_{sn}^j] \begin{bmatrix} \epsilon_{11} & \cdots & \epsilon_{1n} \\ \epsilon_{21} & \cdots & \epsilon_{2n} \\ \vdots & \ddots & \vdots \\ \epsilon_{n1} & \cdots & \epsilon_{nn} \end{bmatrix} \begin{bmatrix} v_{s1}^j \\ v_{s2}^j \\ \vdots \\ v_{sn}^j \end{bmatrix} \quad \text{Eq4}$$

KD-tree with minimum radius search is used to find the blocks that contribute to the estimation of excavated material at location x. It can be observed that the available bucket dig locations includes a significant number of repeatable measurements in the z direction, because the excavator, in general, cuts the bench vertically in a comparatively narrow working area across the bench. Therefore, this excavation process results high uncertainty in vertical direction during excavation and thus all blocks across the bench in the bucket movement direction are forced to be included in the adjacent blocks' list for a given bucket dig position x. Then the random bucket dig locations $x_j$ are simulated on the 3D spatial grid and the space near the given location is convolved using a separable Gaussians coming from each bucket at simulated random locations. KD-tree is again used to find the intersecting blocks to the simulated bucket dig location $x_j$. Spherical shape is assumed for bucket volumes with constant radius. The Figure 2 is showing the schematic representation of the estimation made for a single bucket.

| Realization of bucket dig location at random locations | Pdf of MT distribution at each location | Calculation of pdf | GMM of entire space around a given dig location sampled at random location |

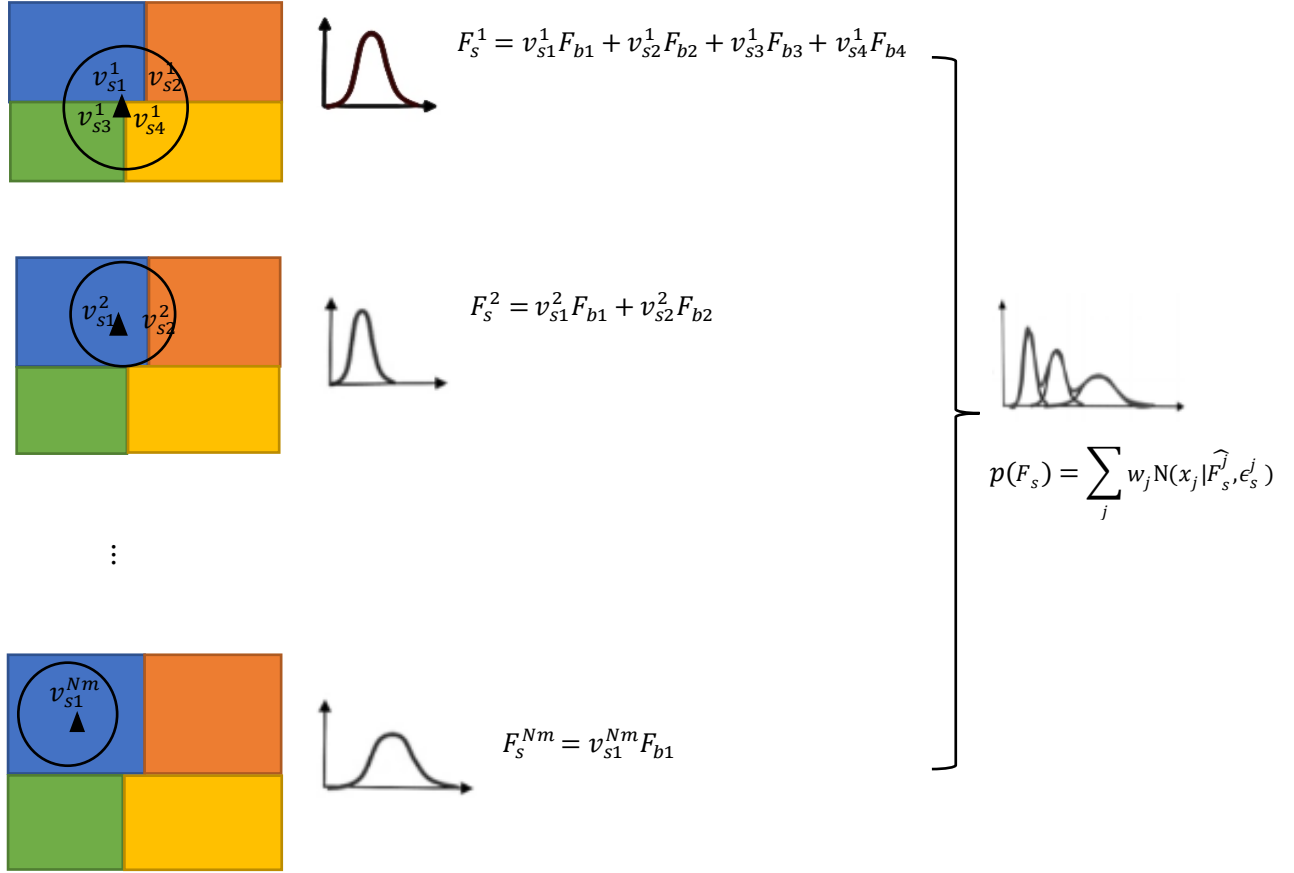

*Figure 2 The estimated grade distribution at each location that is sampled around the given bucket-dig location is integrated into a single GMM. b and s represent block and bucket(scoop) respectively. Coloured square regions show the blocks in vertical direction. Circle region shows the bucket volume that intersects with blocks and the triangle shows the centre of the bucket. Variables in the formulas are as described in the text.*

### 3.1. Aggregating uncertainty from random simulations: GMM and moment matching

GMM is a probabilistic model for representing a mixture of populations with prior Gaussian distribution functions. Mixture distributions provide a useful way for describing heterogeneity in a population, especially when an outcome is a composite response from multiple sources. This paper proposes using GMM to aggregate the conditional distribution of the excavated material independently sampled at random locations.

In our model, suppose Nm bucket dig locations are sampled from the given region. Then the mean and variance, $\{\hat{F}_s^j, \epsilon_s^j\}$, of each excavated material, Figure 2, are calculated using the Eq3 and Eq4. From the bucket volumes at random locations, the joint distribution is represented by the ensemble, $\{(x_1, N(\hat{F}_s^1, \epsilon_s^1)), \ldots, (x_{Nm}, N(\hat{F}_s^{Nm}, \epsilon_s^{Nm}))\}$. A marginal estimate p(Fs) then is the GMM,

$$p(\boldsymbol{F_s}) = \sum_j w_j \text{N}(x_j | \hat{F}_s^j, \epsilon_s^j) \qquad \text{Eq7}$$

composed of Nm equally weighted Gaussian components from the Nm simulated bucket dig locations. The final pdf reflects the spread in the ensemble of conditional realizations used to characterize the uncertainty associated at buckets' content from the given region.

As described in Bar-Shalom et al., 2001, page 56, moment matching is a form of density estimation that is used to reduce a total number of components in a GMM while remaining invariant in the original first and second moment. Thus, the marginal estimation p(Fs) described by Eq7 can be replaced into a single Gaussian with the moments

$$\widehat{\boldsymbol{F_s}} = \sum_j w_j F_s^j$$

$$\boldsymbol{\epsilon_s} = \sum_j w_j \left( \epsilon_s^j + (\hat{F}_s^j - \widehat{\boldsymbol{F_s}})(\hat{F}_s^j - \widehat{\boldsymbol{F_s}})^T \right) \qquad \text{Eq8}$$

The steps used to estimate the moments of a given bucket locations is described in Algorithm 1.

---

**Algorithm 1**: GMM-Moment matching in bucket uncertainty estimation
**Inputs:** Blocks moments $p(F_{b1}, F_{b21}, \ldots, F_{bn})$ around a given bucket dig location, block covariances $\Sigma$, grid locations P(x) within the bench where the bucket dig location falls, bucket volume Vs
**Outputs:** Moments of bucket content ($\boldsymbol{Fs}, \boldsymbol{\sigma s}$) for a given dig location.
**Step1:** Radius neighbour search in three-dimensional on uniform grid locations P(x) = {x1, …, xN},
1. N (q, R) = {p ∈ P | ||p − q|| < R} inside an R of a given query point (bucket dig location) q; R~bench height

**Step2:** Estimate the moments at each grid locations(x) chosen from Step1.
2.   **for** j ← 1 to N **do**
3.     **for** i ← 1 to nj **do**
4.       $vs(i) = \text{Vol}(Vs \cap Fbi)/Vs$     intersecting blocks $F_{bi}$, i=1: nj at sampled location xj
5.       $Fs(j) = Fs(j) + (vs(i) \cdot Fb(i))$
6.     **end for**
7.     $\epsilon s(j) = V\Sigma V^T$               $\Sigma$ is the block covariance, V=[vs(1), vs(2),…,vs(nj)]
8.   **end for**

**Step3:** Estimate the moments of excavated material at given location

9.   **for** j ← 1 to N **do**
10.     $wj = 1/N$
11.     $\boldsymbol{Fs} = \boldsymbol{Fs} + wjFs(j)$
12.     $\epsilon s = \epsilon s + wj(\epsilon s(j) + (Fs(j) - \boldsymbol{Fs})(Fs(j) - \boldsymbol{Fs})^T)$
13.   **end for**
14.   $\boldsymbol{\sigma s} = sqrt(\epsilon s)$

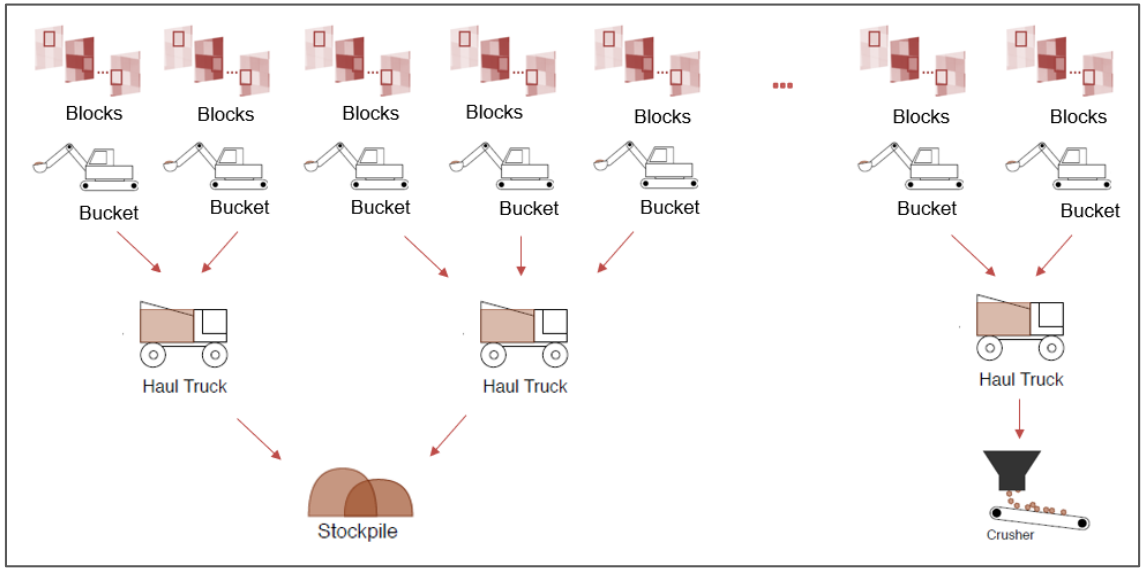

*Figure 3 Work flow of materials transport in an open-pit mine*

In addition to that, this model can be directly implemented to find the uncertainty of the Fe wt% of the material loaded to truck. When the truck is loaded from the near buckets dig locations the values in the buckets are spatially correlated. From the prior conditional probability $p(F_s^j|x_j)$ of Fe wt% of the bucket at given bucket dig locations $x_j$; j=1:n, we can calculate the joint probability distribution $P(F_D|x)$ at dump locations. Here, $P(F_D)$ can be the distribution of Fe wt% of the truck that are directly loaded by buckets coming from N (n=N) dig locations or it can be the joint distribution of buckets coming from M (n=M) dig locations that are sent to a dump location such as stockpiles or crushers; M>N. The overall work flow of materials transport in an open-pit mine can be seen in Figure 3.

The modified Eq1 is given by,

$$F_D = vtF_{s1} + vtF_{s2} + \cdots + vtF_{sn}$$

Where $F_D$ is the estimated value for a dump location(truck/stockpile/crusher) from the estimated buckets' values $F_{si}$; i=1..n for n buckets and $vt$ is the fraction of bucket volume out of total volume transferred. Thus, $vt$=1/n.

If $(F_{s1}, F_{s2}, .., F_{sn})$ are correlated then the joint distribution $p(F_{s1}, F_{s2}, .., F_{sn})$ with moments,

$$\begin{bmatrix} \widehat{F}_{s1} \\ \widehat{F}_{s2} \\ \vdots \\ \widehat{F}_{sn} \end{bmatrix}, \begin{bmatrix} \epsilon_{11} & \cdots & \epsilon_{1n} \\ \epsilon_{21} & \cdots & \epsilon_{2n} \\ \vdots & \ddots & \vdots \\ \epsilon_{n1} & \cdots & \epsilon_{nn} \end{bmatrix}$$

Where $\widehat{F_{si}}$ are the mean of the estimated distribution of bucket content i and $\epsilon_{ij}$ are the covariance between bucket dig locations i and j.

Then the mean of loaded material to truck or dump location is given by

$$\widehat{F_D} = [vt, vt, \ldots, vt]\begin{bmatrix}\widehat{F}_{s1}\\ \widehat{F}_{s2}\\ \vdots\\ \widehat{F}_{sn}\end{bmatrix}$$

Since the buckets' contents are correlated, the covariance as

$$\epsilon_D = [vt, vt, \ldots, vt]\begin{bmatrix}\epsilon_{s11} & \cdots & \epsilon_{s1n}\\ \epsilon_{s21} & \cdots & \epsilon_{s2n}\\ \vdots & \ddots & \vdots\\ \epsilon_{sn1} & \cdots & \epsilon_{snn}\end{bmatrix}\begin{bmatrix}vt\\ vt\\ \vdots\\ vt\end{bmatrix}$$

However, because we are using the volumetric covariance of blocks coming from ore body block model as the priors for the proposed model, the model approach has been further modified to estimates the values in the dump locations directly from the simulated bucket dig locations by generating samples $x_j \sim p(x)$.

Below example shows how the truck values are simulated from the simulated bucket dig locations.

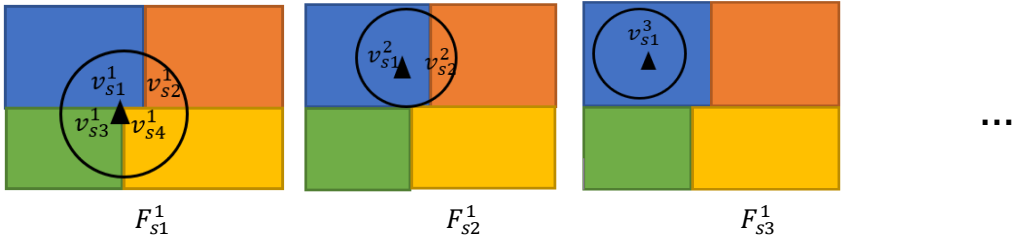

Figure 4 Schematic representation of randomly chosen single simulated bucket dig location for the given dig locations for bucket1, bucket2, bucket 3 and …. $F_{s1}^1, F_{s2}^1, F_{s3}^1$ are the estimated values at simulated locations from the volume intersection model using four blocks, two blocks and a single block respectively. $v_{si}^j > 0$; is simulated bucket for the given dig location j intersecting volumes with blocks i.

If buckets are loaded onto truck T from the recorded bucket dig locations i and we choose the single simulated bucket dig location j given each bucket dig location i, then the values $F_{si}^j$ for those buckets can be estimated using Eq1 (Fig 4).

From Eq1,

estimated value at the simulated location j=1 for bucket 1,

$$F_{s1}^1 = v_{s1}^1 F_{b1} + v_{s2}^1 F_{b2} + v_{s3}^1 F_{b3} + v_{s4}^1 F_{b4}$$

estimated value at the simulated location j=1 for bucket 2,

$$F_{s2}^1 = v_{s1}^2 F_{b1} + v_{s2}^2 F_{b2}$$

estimated value at the simulated location j=1 for bucket 3,

$$F_{s3}^1 = v_{s1}^3 F_{b1}$$

$$\vdots$$

Hence the first simulated truck value is

$$F_T^1 = F_{s1}^1 + F_{s2}^1 + F_{s3}^1 + \cdots$$

$$F_T^1 = v_{s1}^1 F_{b1} + v_{s2}^1 F_{b2} + v_{s3}^1 F_{b3} + v_{s4}^1 F_{b4} + v_{s1}^2 F_{b1} + v_{s2}^2 F_{b2} + v_{s1}^3 F_{b1} + \cdots$$

$$F_T^1 = (v_{s1}^1 + v_{s1}^2 + v_{s1}^3 + \cdots)F_{b1} + (v_{s2}^1 + v_{s2}^2 + \cdots)F_{b2} + (v_{s3}^1 + \cdots)F_{b3} + (v_{s4}^1 + \cdots)F_{b4} + \cdots$$

$$F_T^1 = v_{t1}F_{b1} + v_{t2}F_{b2} + v_{t3}F_{b3} + v_{t4}F_{b4} + \cdots$$

$F_T^1$ is a single Gaussian and $F_{b1}, F_{b2}, F_{b3}$ and $F_{b4}$ are correlated blocks. $v_{si}^j$ is the j$^{th}$ bucket volume intersecting with block i and $v_{ti}$ is the truck volume intersecting with block i; $v_{ti} = \sum_j v_{si}^j$.

The mean is calculated from the estimated bucket values and the variance is estimated using the covariance of the relevant block's using the obk sub block model covariance function.

Similarly, truck values $F_T^i$; i=1: n are simulated using 2$^{nd}$, 3$^{rd}$, ... and N$^{th}$ simulated bucket dig locations of the given dig location 2, 3, ... and n.

$$F_T^2 = F_{s1}^2 + F_{s2}^2 + \cdots + F_{sN}^2$$

$$F_T^3 = F_{s1}^3 + F_{s2}^3 + \cdots + F_{sN}^3$$

$$\vdots$$

$$F_T^n = F_{s1}^n + F_{s2}^n + \cdots + F_{sN}^n$$

The final estimation $F_T$ is calculated from the n simulated truck values $F_T^1, F_T^2, \ldots, F_T^n$ for truck T using moment matching.

The step by step guide to estimate the moments at truck and dump locations are given below.

**Step 1**: For a given truck T (or dump location) find the corresponding bucket dig locations, say N buckets.

**Step2**: For i=1 to N

$A[i] \leftarrow$ the adjacent simulated bucket dig locations at grids. Size(A[i]) =ni

**Step3**: For j=1 to M; where M=min (n1, n2, ..., nN)

$$\widehat{F}_T^j = [vt, vt, \ldots, vt]\begin{bmatrix} \widehat{F}_{s1} \\ \widehat{F}_{s2} \\ \vdots \\ \widehat{F}_{sn} \end{bmatrix}$$

$$\epsilon_T^j = [v_{t1}, v_{t2}, \ldots, v_{tn}]\begin{bmatrix} \epsilon_{s11} & \cdots & \epsilon_{s1n} \\ \epsilon_{s21} & \cdots & \epsilon_{s2n} \\ \vdots & \ddots & \vdots \\ \epsilon_{sn1} & \cdots & \epsilon_{snn} \end{bmatrix}\begin{bmatrix} v_{t1} \\ v_{t2} \\ \vdots \\ v_{tn} \end{bmatrix}$$

Where $\widehat{F}_{s1}^j$ are the mean of the bucket at location j and $\epsilon_{ij}$ are the covariance between the blocks i and j. If the truck volume is estimated from n blocks, then the weight matrix consists a covariance matrix of size n×n. $vt$ is the truck volume divided by number of buckets. $v_{ti}$ is the truck volume intersecting with block i; $v_{ti} = \sum_j v_{si}^j$

**Step4**: Estimate the mean and variance for the truck T

$$\widehat{F_T} = \sum_j w_j \widehat{F}_T^j$$

$$\epsilon_T = \sum_j w_j \left( \epsilon_T^j + (\hat{F}_T^j - \widehat{F_T})(\hat{F}_T^j - \widehat{F_T})^T \right)$$

where $w_j$ is $1/N$.

## 4. Data

Data were collected from the Pilbara iron ore deposit situated in the Brockman Iron Formation of the Hamersley Province, Western Australia. The models were tested on 3477 measured bucket dig locations that were used to excavate a grade block. The blast hole length weighted averaged Fe wt% of grade block is 59.93. Figure5(a) shows the bucket dig locations corresponding to the high-grade grade block that are queried from the load haul cycle data. The green area shows the region of low-grade adjacent block that is located next to the high-grade block. The grade block is situated in the bench X10 and the bench is 10m in height. In the study region, altogether there are 3336 blast holes including the blast holes at above and below benches and the area is further subdivided into sub blocks. There are 684018 sub blocks.

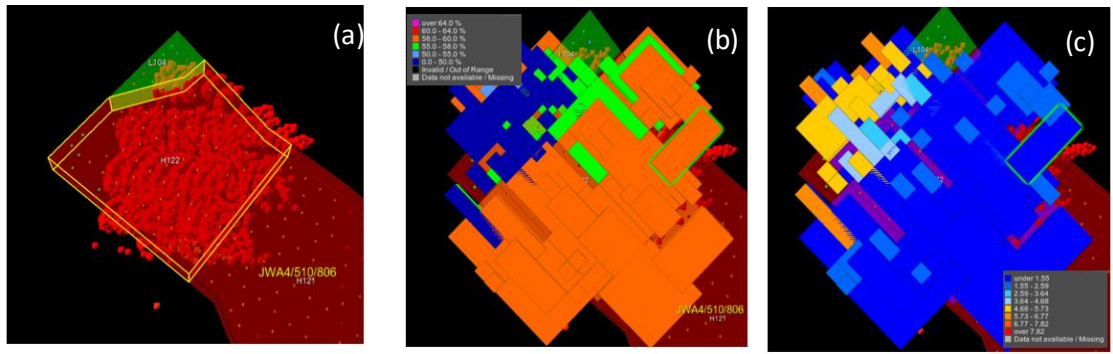

*Figure 5(a) shows the plan view of interested region. The region enclosed by the yellow bounding box shows a high-grade block. The dense red points show the bucket dig locations digging from south to north. (b) shows the plan view of obk sub block model. The colors of the blocks show the estimated mean Fe wt%. Similarly, the colours in the (c) represents the uncertainty (standard deviation) of the estimated Fe wt% of the corresponding sub blocks*

Figure5(b) and (c) show the selected plan view of obk sub blocks that are coloured by their estimated mean Fe wt% and the uncertainty of the estimation respectively. These stats are inferred using the Gaussian process (GP) model that is used to model the spatial stochastic process. As a supervised learning process, GP hyperparameters such as length scales in x, y and z directions, amplitude and noise were learnt using the exploration-only data. The hyperparameters learnt with the corresponding data were then used to inference the mean Fe wt% and the estimation uncertainty of the sub blocks. Estimation of grade distribution of the sub blocks are out of scope of this paper. The proposed model in this paper used the prior information coming from obk sub block models: estimated mean Fe wt%, uncertainty of the estimation and the block covariance matrices.

## 5. Results and Discussion

**Uncertainty of Fe wt% in buckets**

The model was tested on two different test regions chosen from two different benches and the results are presented in Figures 6 and 8. Uncertainty of the prior estimation of adjacent blocks were used in this analysis.

From our models it can be clearly seen that the uncertainty of the estimation in the bucket content is relatively large where the adjacent blocks prior estimated mean values are highly deviate. In figure 6c the circled area shows the area where low and high grades blocks intersect. The main advantage of this kind of probabilistic model is, they clearly highlight us, that the material taken from the circled areas in Figures 6b and 8b are high risk while the material taken from the rest of the area is low risk because we are high confidence of the estimated value of the excavated material from the area (Fig6c and 8c). Knowing this kind of variance is quite important because in the downstream, high uncertain values should be treated with care.

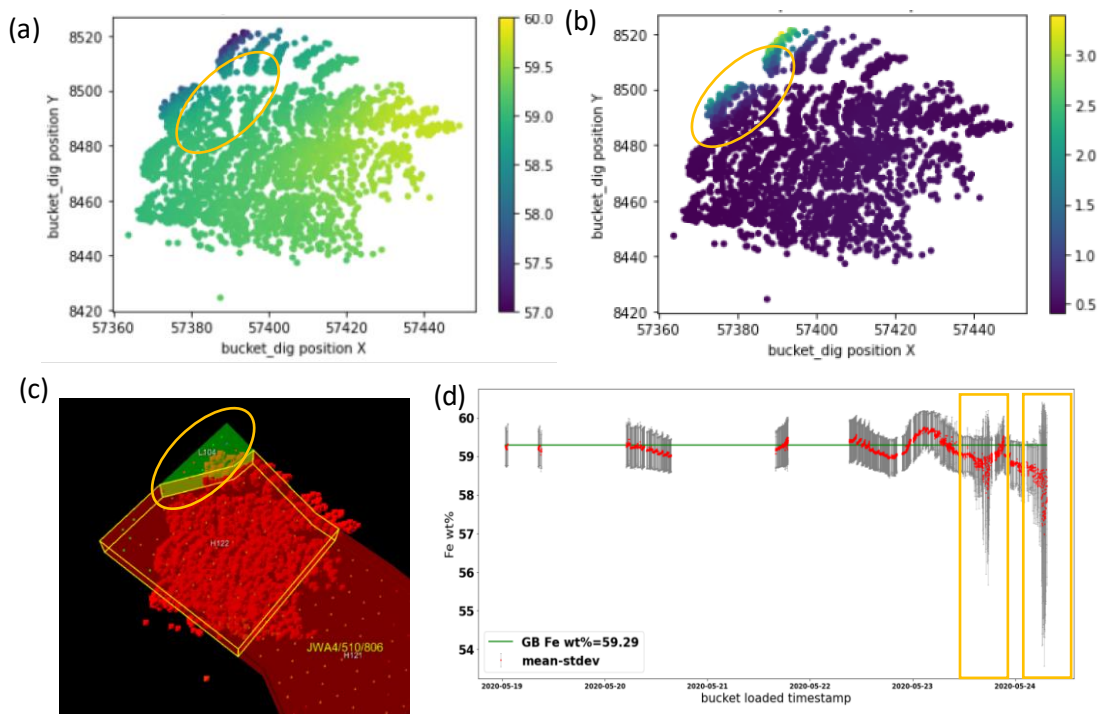

*Figure 6: (a) the estimated mean of the excavated material (b) shows the standard deviation calculated for the given bucket location (c) grade block with the bucket dig location (d) shows the mean and standard deviation of the grade value estimates for each bucket with respect to time of dig.*

The variance calculated by the moment matching does a good job as it is not only capturing the variance of excavated material at each random location, also the mean difference of those pdfs. Figure 7 compares the uncertainty estimation at two bucket locations chosen from low and high-risk areas. Figure 7c clearly shows that the prior mean of the blocks near block boundaries are significantly varied. As seen in Figures 7a and 7b the variance values are achieved when the merged components are close in a normalized distances and Figure 7c and 7d show when the increasing distance between components forces reduction in component variances. The estimated standard deviation for the buckets in Figures 7(a) and 7(c) are 0.51 and 3.08 respectively.

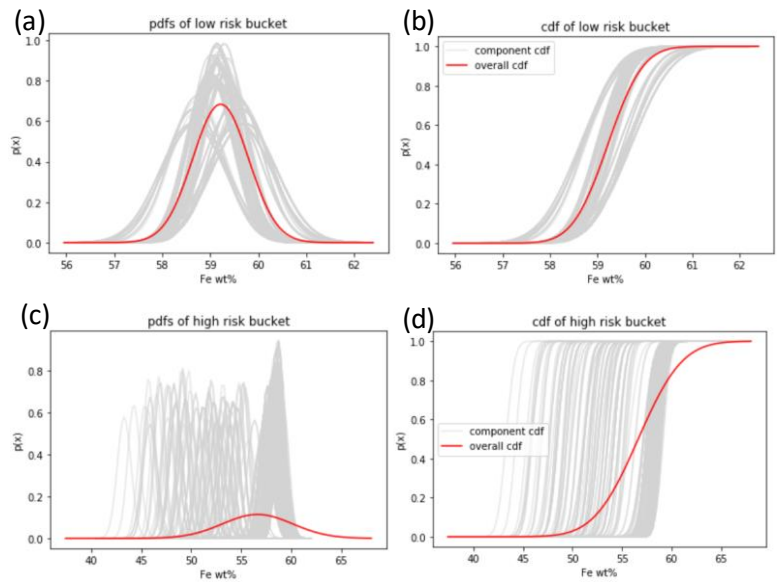

*Figure 7(a) shows the pdfs of the excavated material sampling around the high confident bucket dig location and the red line shows the moments matched from the components. (c) shows the pdfs of the excavated material sampling around the low confident bucket dig location and its corresponding moments matched pdf. (b) and (d) show the cumulative density function of (a) and (c) respectively.*

We run the model in another test region at bench X20 and observed similar results (Fig 8).

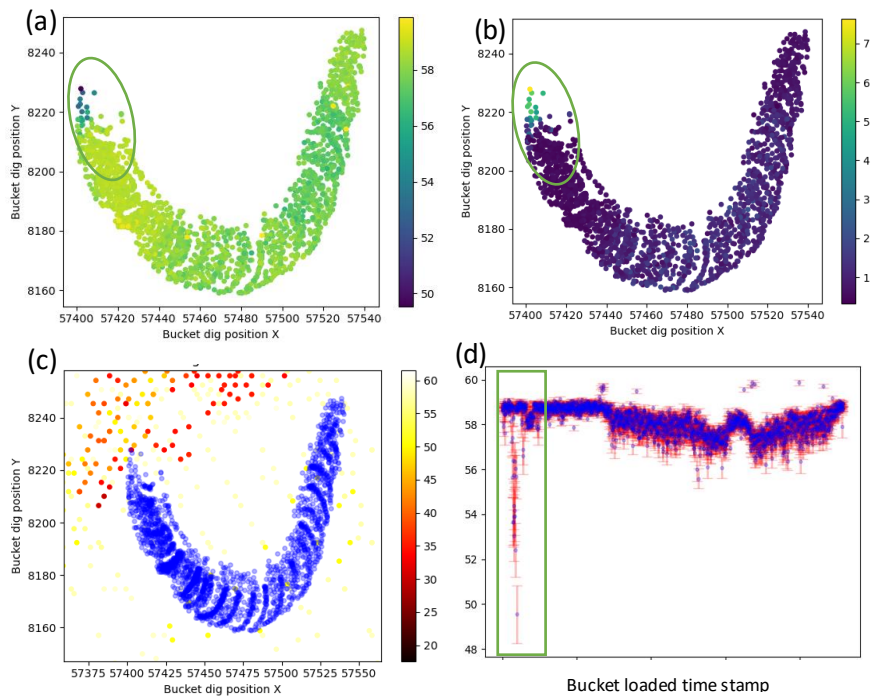

*Figure 8 shows the results obtained at test region 2. (a) and (b) shows the estimated mean and standard deviation of Fe wt% of each bucket material. (c) shows the bucket dig location around the blocks center locations. Block centre locations are coloured according to its bocks prior Fe wt%. (d) shows the moment calculated for each buckets.*

The method proposed in this paper was built on the prior block model that are given with estimated Fe wt% and the uncertainty for each blocks. Grade estimation of these block models was inferred using the ordinary GP and the hyperparameters were learned for each individual geological domain using the assay data coming from exploration drill holes. Based on different interpolation or extrapolation methods to infer the values and different sampling techniques, the estimated moments differ for each block models. Hence, the bucket estimation calculated in this work may not provide the similar results with other 3D block models.

This work has addressed the uncertainty in the excavated material by aggregating the stats provided by the 3D block models. The adjacent blocks are chosen across the bench and 12m distance in XY direction. Figure 9 shows some results obtained by changing some of the parameters used in the proposed model. The results compared estimated Fe wt% and the uncertainty of two buckets that dug at low risk region and the high risk (boundaries at high- and low-grade regions) region. As seen in Figure 9(a), the estimated uncertainty is stable at low risk region and as expected it is varying at high risk region with increasing search radius. Increasing the search radius includes more grid locations that are sampled around the given bucket locations, hence increase the number of components contributed in final estimation. On the other hand, increasing the sample interval in Figure 9b, includes a smaller number of Gaussian components in the final calculation. Sampling interval on grids doesn't impact the results obtained at low risk region but at high risk region (Fig9(b)).

The model was further tested with different bucket sizes provided in different mining excavators' manuals. The volumes of the buckets change from 15 m$^3$ to 45 m$^3$. The estimated uncertainty stays stable for different volumes of buckets that dug at both low and high-risk regions (Fig 9c).

The model can be further modified by weighting the sampled bucket dig locations for a given bucket dig location using the inverse distance squared, thus the uncertainty coming from the closest buckets' moments are more likely to contribute on the estimation for GMM. The model is flexible to adjust the search parameters.

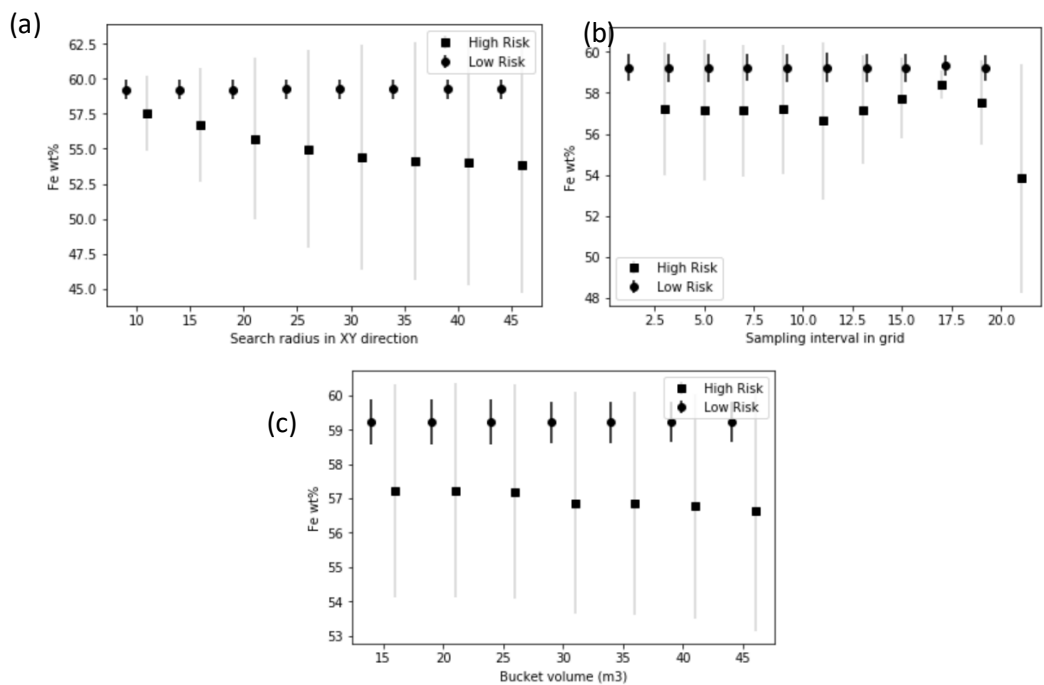

*Figure 9 Estimated mean and standard deviation versus the (a)search radius, (b)sampling interval on grids and (c) different bucket volumes.*

## Propagation of uncertainty via buckets

Using the 3477 buckets analysed in this study region, there were altogether 342 trucks loaded and then the trucks dumped at ten locations including various stock piles and crushers. The results are presented in Figure 10.

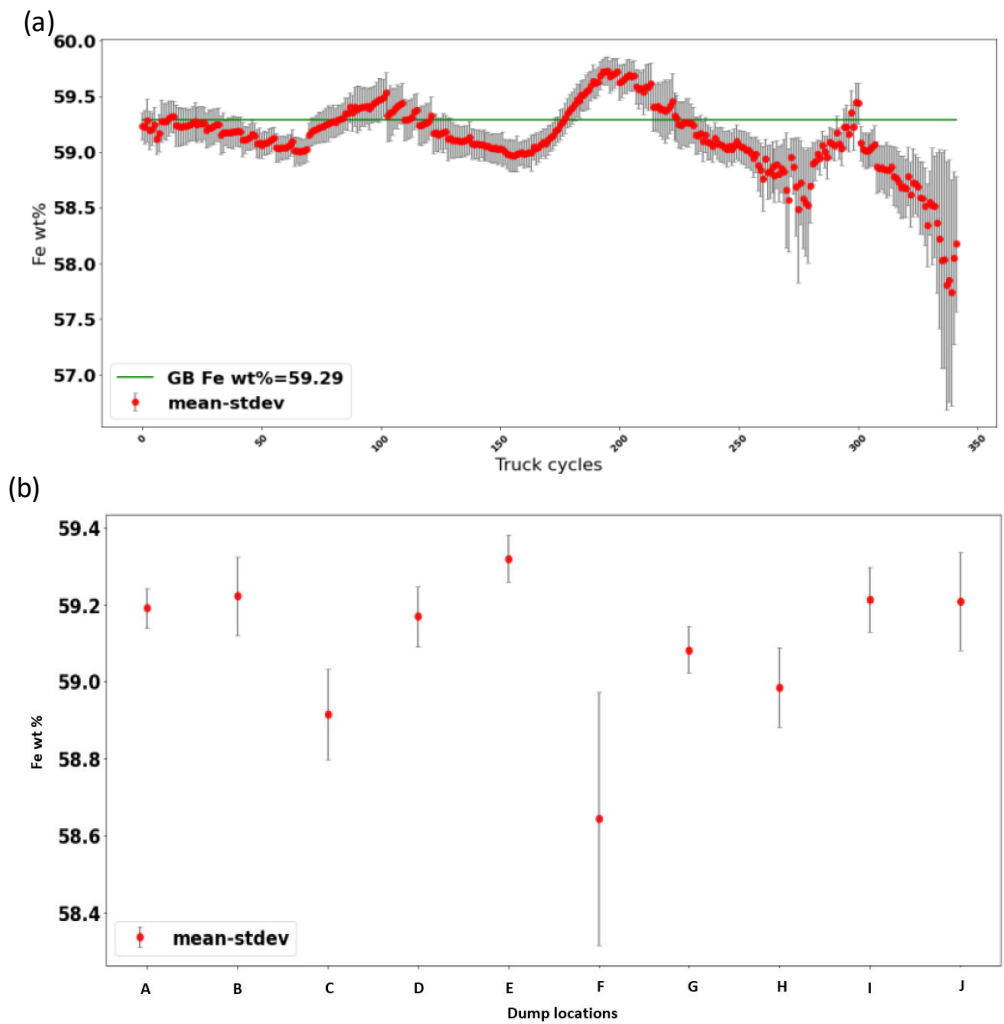

*Figure 11 shows the estimated moments of Fe wt% at (a) 342 trucks and (b) ten dump locations.*

As expected, the uncertainty estimated in the buckets reflects in the material loaded to truck. Hence high uncertainty is observed in the trucks that carried the material from the buckets near high risk regions (Fig10a).

The data used in this paper were collected from a single grade block, thus the material transported to dump locations are spatially correlated. The same proposed method that used to infer the truck material was used for estimating the mean and uncertainty of Fe wt% at dump locations (Fig10b). In order to infer the moments at the dump locations, the model assumed that the trucks are continuously arrived at the dump locations. Hence the model aggregates the continuous truck loads to the destination and the same spatial correlation applied. As seen in Figure 10b, the high uncertainty at the dump location F is due to the high uncertain truck loads that were loaded near high risk region. However, the uncertainty aggregated from the grade blocks are continuously significantly decreasing at trucks and then at dump locations (Fig 11). This is because, that the uncertainty of Fe wt% for a given bucket is estimated from the random sample locations, say n, and then the uncertainty of Fe wt % for a given truck is estimated from multiple buckets, say m, thus the samples are coming from nm simulated locations. Hence, increasing sample size reduces the standard deviation, but the mean doesn't change significantly.

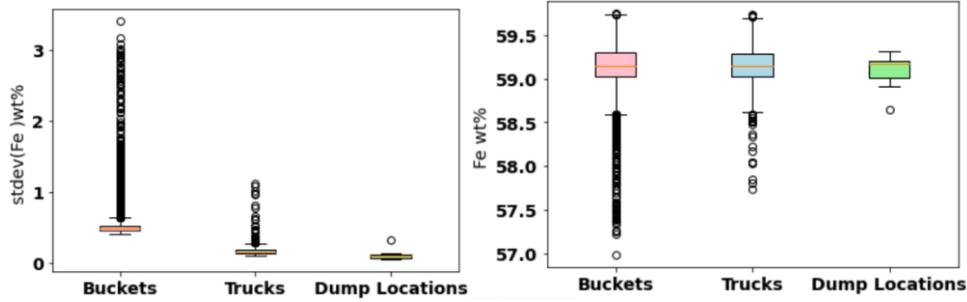

*Figure 11 compares the overall mean and standard deviation of Fe wt% estimated for buckets, trucks and dump locations.*

However, this might not be the case in reality. At the dump locations, different trucks coming from different load locations can dump materials sequentially. Hence the aggregated material for a given time period, doesn't need to be coming from the same load locations. That is, because the trucks are sent to different dump locations, usually trucks at dump location for a given period are possibly coming from the distance bigger than the variogram range, thus the assumption of independency is hold (Neves, 2020). So, the trucks dump at this destination could be loaded from spatially uncorrelated regions. Once we know the distribution of Fe wt% in trucks from the model provided above, the distribution of the grade over a certain time period arrived at the stockpile or plant can be directly inferred using moment matching using the Eq8. Where the moments ($\widehat{F_D}, \epsilon_D$) of Fe wt% of material dump at given dump location for a given period Δt is inferred from the prior component moments ($\hat{F}_T^j, \epsilon_T^j$) of Fe wt% of N trucks arrives at the dump location during Δt, j=1:N and $w_j = 1/N$.

$$\widehat{F_D} = \sum_j w_j \hat{F}_T^j$$

$$\epsilon_D = \sum_j w_j \left( \epsilon_T^j + (\hat{F}_T^j - \widehat{F_D})(\hat{F}_T^j - \widehat{F_D})^T \right)$$

The method proposed for bucket estimation in this paper is also beneficial to infer the moments of the material in the trucks, thus at dump locations, when we don't have the actual bucket dig positions. Sometimes, bucket sensors frequently go to off nominal states and the bucket dig positions are not always recorded for every truck load. However, inferencing the representative loading position in the bench is possible from the other sensor measurements coming from the excavators (Balamurali, 2021). Once this position is known for the truck, the estimated mean and standard deviation of the material to truck can be obtained from the blocks across the bench for a certain radius in XY direction. The estimations could be reasonably accurate in the low risk regions.

However, high uncertainty can be expected at the borders where grades are varied significantly. This is subject to further investigation.

The major pitfall of this model is, the model doesn't capture the dynamic of volume change as the materials are removed from the ground with every bucket dig location. Thus, extending the current model as a dynamic model, could be a possible solution for this particular problem. In addition to this, ore dilution and loss can occur with blasting and sheeting, grading of roads and benches, that moves the blocks from their original place. These issues have been further investigated in the continuation of this study.

## 6. Conclusion

A method is proposed to estimate the uncertainty in excavated material by aggregating the prior blocks' moments. The model implementation is reasonably simple if we know the parametric description of each block or point locations. No studies in literature utilise block models to infer uncertainty attach to the material taken into bucket volumes and thus onto truck and other destinations, and no approaches use diggers sensor data as model inputs for this purpose. Hence the identified gap in research with which this paper addresses.

From our models it can be clearly seen that the uncertainty of the estimation in excavated material is relatively large where the adjacent blocks' prior mean values are highly deviate. Our model approach has proved that by probabilistically model the material at different locations such as bucket, truck and stockpile, allows us to identify the high-risk material which needs further processing down the tracking pipeline.

The model will be continuously improved to address the pitfalls of the current model.

**Acknowledgment**

This work was supported by the Australian Centre for Field Robotics and the Rio Tinto Centre for Mine Automation. The author would like to thank Tim Bailey for his invaluable insights and Alexander Lowe and OBK technical team as their contributions to the software implementation for nearest block search and block covariance estimation have allowed the proposed models to be experimented and evaluated on. The author extends her gratitude to above ground model team within RTCMA and RIO who have involved in frequent discussions.

**References**

1. Balamurali, M. (2021), Inferencing the earth moving equipment-environment interaction in open pit mining, Open Pit Operators Conference, AUSIMM (abstract accepted; paper in preparation)

2. Bar-Shalom, Y., Li, X., and Kirubarajan, T. (2001). Estimation with Applications to Tracking and Navigation. John Wiley and Sons.
3. Boucher, A. and Dimitrakopoulos, R. (2012) Multivariate Block-Support Simulation of the Yandi Iron Ore Deposit, Western Australia. Math Geosci 44, 449–468 https://doi.org/10.1007/s11004-012-9402-9
4. Dimitrakopoulos, R.; Godoy, M. (2014) Grade control based on economic ore/waste classification functions and stochastic simulations: Examples, comparisons and applications. Min. Technol., 123, 90–106
5. Gilardi, N., Bengio, S., and Kanevski, M. (2002) Conditional Gaussian Mixture Models for Environmental Risk Mapping. InNNSP, URLhttp://www.idiap.ch.
6. Isaaks EH, Srivastava RM (1989) An introduction to applied geostatistics. Oxford University Press, New York, p 561
7. Isaaks E, Treloar I, Elenbaas T (2014) Optimum dig lines for open pit grade control. In: Proceedings of the 9th international mining geology conference, Australian institute of mining and metallurgy, pp 425 – 432
8. Melkumyan, A. and Ramos, F. (2009) A Sparse Covariance Function for Exact Gaussian Process Inference in Large Datasets, Proceedings of the International Joint Conferences on Artificial Intelligence Organization, vol. 9, pp. 1936-1942.
9. Neves, J., Araújo, C. & Soares, A. Uncertainty Integration in Dynamic Mining Reserves. Math Geosci (2020). https://doi.org/10.1007/s11004-020-09866-1
10. Norrena K, Deutsch C (2001) Automatic determination of optimal dig limits accounting for uncertainty and equipment constraints. In: Centre for computational geostatistics Report 3–114
11. Qu, J., & Deutsch, C. (2017). Geostatistical Simulation with a Trend Using Gaussian Mixture Models. Natural Resources Research, 27, 347-363.
12. Rossi ME, Deutsch CV (2014) Mineral resource estimation. Springer, Dordrecht, p 372
13. Silva DSF, Deutsch CV (2017) Multiple imputation framework for data assignment in truncated pluri-Gaussian simulation. StochEnviron Res Risk Assess 31(9):2251–263.https://doi.org/10.1007/s00477-016-1309-4
14. Vasylchuk, Y. V., & Deutsch, C. V. (2015). A short note on truck-by-truck selection versus polygon grade control. In Proceedings of the 17th CCG Annual Conference, (paper 312). Edmonton, AB, Canada: University of Alberta.
15. Verly G. (2005) Grade control classification of ore and waste: a critical review of estimation and simulation based procedures. Math Geol. 37(5):451–475. doi:10.1007/s11004-005-6660-910.1007/s11004-005-6660-9
16. Wilde, B. J., & Deutsch, C. V. (2007). A short note comparing Feasibility Grade Control with Dig Limit Grade Control. In Proceedings of the 9th CCG Annual Conference, (paper 302). Edmonton, AB, Canada: University of Alberta.
17. Williams, J. (2003). Gaussian mixture reduction for tracking multiple manoeuvring targets in clutter. Master's thesis, Air Force Institute of Technology, Wright-Patterson Air Force Base.